\documentclass[conference,10pt]{IEEEtran}
\IEEEoverridecommandlockouts
\usepackage{algorithm}
\usepackage{algorithmic}
\usepackage{amsfonts}
\usepackage{amsmath}
\usepackage{mathrsfs}
\usepackage{color}
\usepackage{esvect}
\usepackage{gensymb}
\usepackage{graphicx}
\usepackage{mathtools}
\usepackage[binary-units=true]{siunitx}
\usepackage{stfloats}
\usepackage{url}
\usepackage{xspace}
\usepackage{wrapfig}
\usepackage[normalem]{ulem}
\usepackage{subcaption}
\usepackage[shortlabels]{enumitem}

\makeatletter
\newcommand\fs@spaceruled{\def\@fs@cfont{\bfseries}\let\@fs@capt\floatc@ruled
  \def\@fs@pre{\vspace{0.5\baselineskip}\hrule height.8pt depth0pt \kern2pt}%
  \def\@fs@post{\kern2pt\hrule\relax\vspace{-0.4cm}}%
  \def\@fs@mid{\kern2pt\hrule\kern2pt}%
  \let\@fs@iftopcapt\iftrue}
\makeatother
\floatstyle{spaceruled}
\restylefloat{algorithm}
\usepackage[font=small]{caption}
\usepackage[sort,square,numbers]{natbib}
\newcommand{\eat}[1]{}

\newcommand{\para}[1]{\smallskip \noindent {\bf #1}}
\newcommand{\softpara}[1]{\smallskip \noindent \underline{#1}}

\newcommand{\showOutline}{s}
\newcommand{\hide}[1]{}

\definecolor{darkgreen}{rgb}{0,0.5,0}
\definecolor{darkblue}{rgb}{0,0,0.5}

\newcommand{\outline}[2]{\if#1\showOutline {\color{darkgreen}#2} \fi}
\newcommand{\detailedoutline}[2]{\if#1\showOutline {\color{darkblue}#2} \fi}

\newcommand{\todo}[2]{\if#1\showOutline {#2} \fi}

\newcommand{\purp}[1]{} 

\def\blackbox{\hfill {\vrule height6pt width6pt depth0pt}}
\newcommand{\vsplem}{\vspace{0.025in}}

\newtheorem{lem}{Lemma}
\newtheorem{thm}{Theorem}
\newtheorem{obv}{Observation}

\newenvironment{lem-wo-prf}{\begin{lem} \nopagebreak}{{\hfill$\blackbox$} \end{lem}}
\newenvironment{lem-w-prf}{\begin{lem} \nopagebreak}{\end{lem}}
\newenvironment{thm-w-prf}{\begin{thm} \nopagebreak}{\end{thm}}
\newenvironment{thm-wo-prf}{\begin{thm} \nopagebreak}{{\hfill$\blackbox$} \end{thm}}

\newcounter{packednmbr}

\newcommand{\dqc}{{\tt DQC}\xspace}

\newcommand{\onlinedqc}{{\tt OS-DQC}\xspace}

\newcommand{\qap}{{\tt QAP}\xspace}

\newtheorem{defin}{Definition}

\newcommand{\GSCH}{\mbox{\tt GreedySC-Heuristic}\xspace}

\newcommand{\bffifo}{\mbox{\bf FIFO}\xspace}
\newcommand{\bfshortest}{\mbox{\bf Shortest}\xspace}
\newcommand{\bfmmft}{\mbox{\bf MMFT}\xspace}
\newcommand{\fifo}{\mbox{\tt FIFO}\xspace}
\newcommand{\shortest}{\mbox{\tt Shortest}\xspace}
\newcommand{\mmft}{\mbox{\tt MMFT}\xspace}
\urldef\paperdoi\url{https://doi.org/10.1109/QCNC69040.2026.00045}
\title{Online Scheduling and Distribution of Quantum Circuits over Quantum Networks%
\thanks{Full author version revised after acceptance. A shortened version,
prepared to meet the conference page limit, was published in
\emph{2026 International Conference on Quantum Communications, Networking, and Computing (QCNC)},
pp.~264--270. DOI: \paperdoi.
\par\textcopyright{} 2026 IEEE. Personal use of this material is permitted.
Permission from IEEE must be obtained for all other uses, in any current or future media,
including reprinting/republishing this material for advertising or promotional purposes,
creating new collective works, for resale or redistribution to servers or lists,
or reuse of any copyrighted component of this work in other works.}}

\author{
\IEEEauthorblockN{Yukun Yang, Himanshu Gupta}
\IEEEauthorblockA{\textit{Department of Computer Science, Stony Brook University, Stony Brook, NY 11790, USA}}
}

\begin{document}
\maketitle
\thispagestyle{plain}
\pagestyle{plain}

\begin{abstract}
Emerging quantum networks enable \emph{distributed quantum computing} (\dqc) by interconnecting modest, noisy quantum processors. However, practical deployments face tight memory budgets, probabilistic entanglement generation, and decoherence, all while serving multiple users. In realistic settings—cloud execution, iterative/interactive workloads and concurrent jobs—circuits do not arrive in a single batch but as a stream. This creates a fundamentally online problem: the system must react to stochastic arrivals, balance latency and throughput, and remain stable under varying loads without foresight of future jobs.

We study \emph{Online Scheduling and Distribution of Quantum Circuits} (\onlinedqc) and formalize the online scheduling problem for quantum networks: circuits arrive over time and must be executed without knowledge of future arrivals. At each \emph{event time} (arrival or completion), the system decides which waiting circuits to launch and how to place their qubits, while accounting for congestion from in-flight circuits and constraints due to finite memories and decoherence. The objective is to minimize the mean completion time and makespan under these physical and resource constraints.

\end{abstract}
\section{\bf Introduction}
\label{sec:intro}


Quantum networks are emerging as the connective tissue for distributed quantum computation: small processors linked by photonic channels can share entanglement and coordinate nonlocal operations across distance. Near-term deployments remain modest and multi-tenant, so workloads look less like coordinated experiments and more like a stream of independently issued jobs. Circuits appear at arbitrary times, vary in size and urgency, and must coexist on scarce quantum memories and limited, unreliable links. Without advance knowledge, the scheduler must decide in real time which circuits to execute and where to place them, maintaining low latency and high throughput while keeping the platform stable as demand fluctuates.

We therefore take an online view: arrivals, contention, and progress coevolve in real time, and performance is governed not only by circuit structure but also by admission and placement choices, coherence-aware batching of on-demand entanglement, and congestion management on the shared links. 

\para{Prior Work.}
The foundational results on quantum repeaters and long-distance entanglement established the substrate for networked quantum computation \cite{Briegel1998QuantumRT}. Vision and systemization work toward a quantum internet that articulates layered designs in which link-level entanglement services support network and application control, with prototypes validating these interfaces \cite{Wehner2018,Dahlberg2019,Pompili2022}. On the network side, the entanglement distribution has been studied under probabilistic links and finite memories, with algorithms that optimize distribution (routing, fusion, purification) and resource usage in the presence of stochastic success \cite{Pant2019,fan2023optimized,fan2024distribution}. These efforts characterize the communication substrate and the constraints that any higher-level scheduler must respect.

In parallel, the \dqc and compilation communities have explored placement, routing, and partitioning across devices, including qubit routing and architecture-aware mapping \cite{Cowtan2019,Zulehner,Siraichi,tket}, compiler and runtime support for multi-device execution \cite{Daei2020OptimizedQC,Caleffi,Ferrari-etal-2021,g2021efficient}, and entanglement-assisted distribution mechanisms over general network topologies \cite{dqc-gen,sundaram2023distributing,dqcqr}. Closest to our setting, a recent batched multi-circuit study considers efficient execution when all jobs are known in advance \cite{Yang2025Efficient}; here we address streaming arrivals and study online admission and execution without lookahead. Classical scheduling offers latency-oriented guidance, for example, the optimality of the shortest remaining processing time in exact sizes, while our service times are shaped by congestion and coherence limits \cite{Schrage1968}. Finally, placement objectives in \dqc naturally connect to quadratic assignment formulations, for which approximation and heuristic results are available \cite{Esther}.

\para{Our Contributions.}  
In this paper, we formulate and address the problem of online scheduling and distribution of quantum circuits. Our
key contributions are as follows.
\begin{itemize}
\item \emph{Event-Driven Online DQC with an Oracle.}
We snapshot the in-flight state, maintain congestion-aware routing, and compute placements on currently free memories. An oracle maps any candidate launch set to predicted per-circuit finishing times and the latest finishing time, keeping the policy layer independent of single-circuit mechanics.

\item \emph{Online Admissions.}
We implement three rules—\fifo; \shortest (earliest predicted finish); \mmft (smallest marginal tail)—and admit one circuit at a time, refreshing after each commit until no fit remains. Our evaluation also compares these with an offline global scheduler (\GSCH).

\item \emph{Unified EP-Multiset Batching.}
We unify unfinished EPs with those induced by newly admitted circuits into a single multiset; partition into sequential batches that respect memory/channel capacities and link concurrency; and order batches so that every EP is consumed within the decoherence threshold.
\end{itemize}

\section{\bf Background}
\label{sec:back}

\begin{figure*}
\vspace{-0.35in}
  \begin{minipage}[t]{.67\linewidth}
    \includegraphics[width=\linewidth]{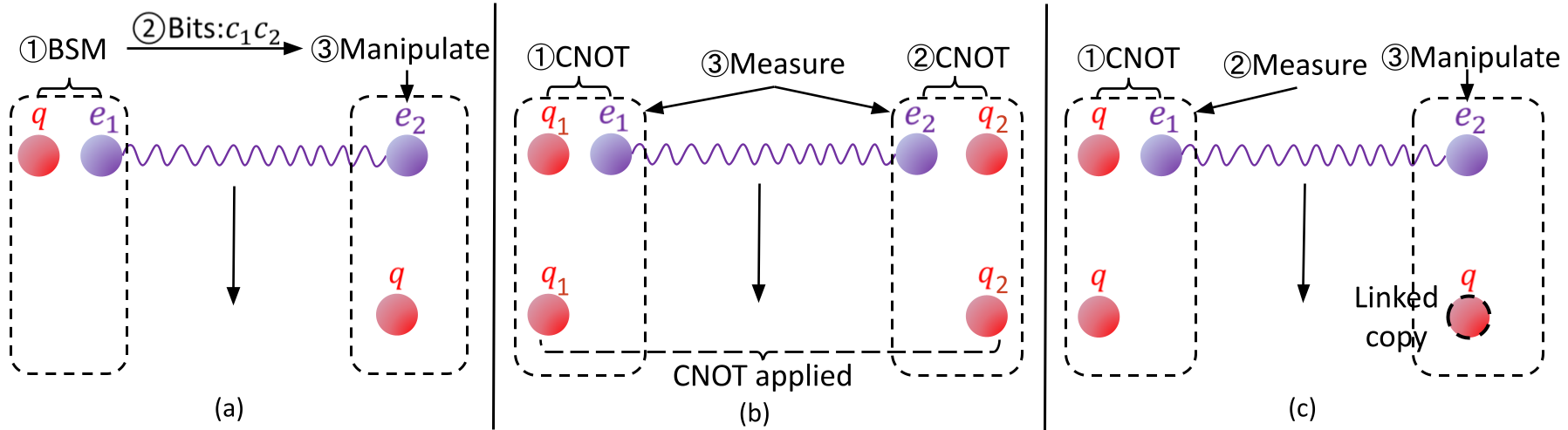}%
    \caption{Quantum communication. (a) Teleportation, (b) Telegate, (c) Cat-Entanglement. Dashed boxes are the network nodes; the initial (final) state of the qubits is at the top (bottom).}%
    \label{fig:Communication}
  \end{minipage}\hfil
  \begin{minipage}[t]{.32\linewidth}
    \includegraphics[width=\textwidth]{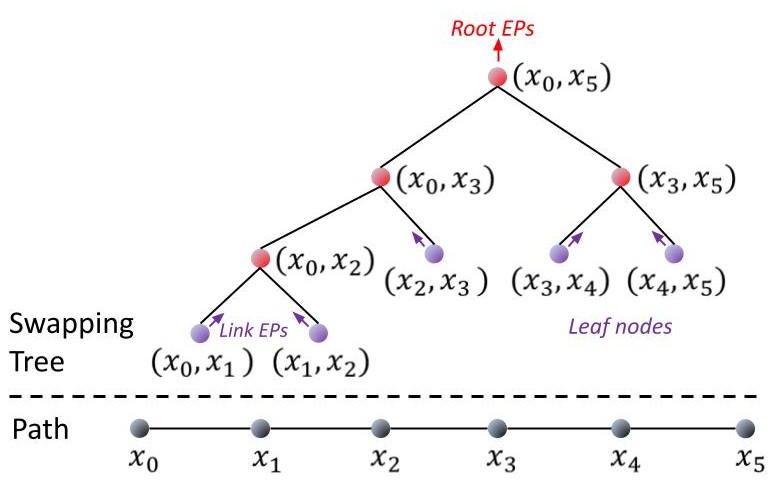}
    \caption{Swapping tree to generate remote EPs.}
    \label{fig:swapping_tree}
  \end{minipage}%
  \vspace{-0.2in}
\end{figure*}

\para{Quantum Circuit and Network.}
We represent a quantum circuit $C$ over a set of qubits $Q=\ \{q_1, q_2, \ldots\}$ as an ordered list of gate tuples: A single-qubit/unary gate on $q_i$ at the sequence index $\ell\in\{1,2,\ldots \}$ is written as $(q_i,\ell)$, and a two-qubit/binary gate on $(q_i,q_j)$ at the sequence index $\ell$ is $(q_i,q_j,\ell)$. The sequence index $\ell$ gives a linear order consistent with dependencies: gates that act on the same logical qubit appear in program order, and any gate that is conditioned on a prior measurement appears after that measurement. The quantum network is a connected graph $G=(V,E)$. Each node $u\in V$ is a quantum computer (QC) with $M_u$ quantum memories. Each edge $e\in E$ denotes a quantum channel between its incident nodes, with an associated classical side channel.

\softpara{Qubit Assignment.}
A placement $\eta$ assigns each circuit qubit to exactly one quantum memory in the network. For a binary gate $(q_i,q_j,\ell)$, let $u$ be the node that hosts $\eta(q_i)$ and $v$ be the node that hosts $\eta(q_j)$. The gate is \emph{local} if $u=v$ and \emph{remote} otherwise.

\para{Remote Gate Primitives.}
For any \emph{remote} gate $(q_i,q_j,\ell)$ whose operands are hosted at nodes $u$ and $v$ under $\mu$, we realize the interaction using one of three standard primitives (see Figs.~\ref{fig:Communication}(a)--(c)).
A maximally-entangled pair (EP) between nodes $u$ and $v$, denoted $\mathrm{EP}(u,v)$, is a shared Bell pair across the endpoints; consuming one EP enables a single instantiation of any primitive below. 

\softpara{Teleportation.}
This primitive relocates a data qubit to co-locate the operands: create $\mathrm{EP}(u,v)$, perform a Bell-state measurement at $u$, send two classical bits to $v$, and apply Pauli corrections so that the logical qubit now resides at $v$ \cite{Bennett+:93}. See Fig.~\ref{fig:Communication}(a). Each teleportation consumes one EP; teleporting the qubit back later requires an additional EP.

\softpara{Telegate.}
This primitive enacts a remote $\mathsf{CNOT}$/$\mathsf{CZ}$ in place without relocating data qubits \cite{Caleffi}: create $\mathrm{EP}(u,v)$, have each side locally couple its data qubit to its half of the EP, then measure the EP halves, and apply the corresponding Pauli corrections on the data qubits to complete the nonlocal gate. See Fig.~\ref{fig:Communication}(b). Each telegate invocation consumes one EP.

\softpara{Cat–Entanglement (Linked Copy).}
This primitive creates a read-only linked copy of a control qubit to one or more target qubits using shared entanglement; subsequent fan-out $\mathsf{CNOT}/\mathsf{CZ}$ gates with the \emph{same} control reuse the link until a single-qubit gate acts on the control or the link expires \cite{Eisert+:00,YimsiriwattanaL:05}. See Fig.~\ref{fig:Communication}(c). Each link setup consumes one EP per control–target pair; reusing the link for additional gates does not consume additional EPs until disentanglement/expiration. When a unary operation needs to be executed on the control qubit, a disentanglement operation is performed, which destroys the linked copies; afterwards, the unary operation is applied.

\softpara{Remote Gate Primitive Selection.}
All three primitives consume EPs: teleportation and telegate use one EP per invocation, while a cat–entanglement uses one EP per control–target link in setup and can be reused across multiple gates. Teleportation transports one operand to the node where follow-up operations can proceed locally—useful when many subsequent gates involve that qubit at the destination. Telegate executes the remote gate in place without relocating data qubits—well suited for one-off interactions or when moving a qubit would later force additional moves. Cat–entanglement amortizes its setup over several two-qubit gates that share the same control; it is particularly natural for $\mathsf{CZ}$-only subcircuits, and the link must be torn down before any subsequent unary gate on the control (after which a fresh setup would be required). \eat{In this work, we enable all three and select the primitive based on the current qubit placement and the surrounding gate pattern.}

\para{Generating EPs over Remote Nodes.}
One simple way to distribute an EP between nodes $u$ and $v$ is to create a Bell pair locally and physically send its two halves to $u$ and $v$; however, direct transport of qubits over long distances is highly lossy and can introduce irrecoverable errors. Instead, for each remote gate between $u$ and $v$, we select a path $p:u\leadsto v$ in $G$ and use \emph{entanglement swapping}: first pump each edge on $p$ to obtain fresh short-range Bell pairs (and, if needed, extra pairs for purification), then perform Bell-state measurements at intermediate nodes to make them into an end-to-end $\mathrm{EP}(u,v)$. For example, Fig.~\ref{fig:swapping_tree} illustrates a swapping tree that generates an EP between $x_0$ and $x_5$ using pairs along the path $x_0, x_1, x_2, x_3, x_4, x_5$. Different swapping trees (orders/groupings of swaps) on the same path can incur different generation latencies~\cite{DPpaper,real-timeqce24}. The selection of swapping trees may be guided either by a baseline shortest-path cost derived from physical link parameters or by a congestion-aware effective path cost that increases with current load.

\softpara{Generation Latency of EPs.}
EP generation is stochastic at the link level and uses shared network resources along $p$, so multiple binary gates may contend when scheduled together. When generating several EPs concurrently, these shared resources must be allocated across requests, which in turn affects completion time.

\softpara{Decoherence and Fidelity of EPs.}
\emph{Fidelity} measures how close a realized state is to the ideal and degrades with storage time and operations; the time-driven component is called \emph{decoherence}. We enforce a \emph{decoherence threshold} $\tau$: an EP created at time $t$ must be consumed by $t{+}\tau$. To counteract low fidelity, we employ \emph{entanglement purification}, which probabilistically distills a higher-fidelity pair from multiple raw EPs at the cost of additional pairs and delay \cite{Briegel1998QuantumRT,fan2024distribution}.

\section{\bf Problem Formulation}
\label{sec:problem}

We start by defining the key terms and models used throughout this paper; with these in place, we then present the formal problem statement and objectives.

\para{Memory Graph and Computers.}
We model the network at the \emph{memory} level using a weighted undirected graph
\[
G_M=(V_M,E_M,w,\mathrm{cap}),
\]
where each $v\in V_M$ is a quantum memory; each $e=(u,v)\in E_M$ is a link on which Bell pairs (EPs) can be generated; $w(e)>0$ is a base per-link cost (e.g., a distance/latency proxy); and $\mathrm{cap}(e)\in\mathbb{Z}_{\ge 1}$ is the number of parallel channels on $e$. Memories belong to quantum computers (QCs); an edge is \emph{remote} if it connects memories on different QCs and \emph{local} otherwise. Throughout, the cost of \emph{local} (intra-QC) gates/paths is negligible compared to the remote generation of EP.

\softpara{Congestion-Aware Effective Distances.}
Running circuits consumes EPs and thereby increases link load. Let $y_e\in\mathbb{Z}_{\ge 0}$ be the number of \emph{unfinished} EPs whose currently planned routes include edge $e$. We inflate the base weight $w(e)$ by a dimensionless congestion parameter $\beta\ge 0$:
\[
w_{\mathrm{eff}}(e)\;=\;w(e)\Bigl(1+\beta\,\frac{y_e}{\mathrm{cap}(e)}\Bigr).
\]
For any memory pair $(u,v)$, the effective distance is the shortest-path metric on these inflated weights,
\[
B_{\mathrm{eff}}(u,v)\;=\;\min_{P\in\mathcal{P}(u,v)}\ \sum_{e\in P} w_{\mathrm{eff}}(e),
\]
where $\mathcal{P}(u,v)$ is the set of simple paths that join $u$ and $v$. $\beta=0$ recovers the base metric $B$, and a larger $\beta$ makes links of high use relatively more expensive. When focusing on \emph{remote} communication, intra-QC edges are given zero weight and carry no load contribution (i.e., $w_{\mathrm{eff}}(e)=0$ and $y_e=0$ for local edges), so only inter-QC paths affect $B_{\mathrm{eff}}$. This distance is then used to locate the highly interacting QC neighbors and to steer new EP routes away from congested links.

\para{Circuits and Flow.}
Let $n$ be the number of circuit qubits, and let $G_2(C)$ be the collection of two-qubit gate instances for the circuit $C$. We construct the symmetric flow matrix $A(C)\in\mathbb{Z}_{\ge 0}^{n\times n}$ as follows:
\begin{enumerate}
\item Initialize $A(C)$ as the zero matrix $n\times n$.
\item For each gate instance $g\in G_2(C)$ acting on qubits $(i,j)$ with $i\neq j$, increase $A_{ij}(C)$ and $A_{ji}(C)$ by one.
\end{enumerate}
By construction, $A$ is symmetric with a zero diagonal and records pairwise interaction counts. Equivalently, for any distinct $i$ and $j$, $A_{ij}(C)$ equals the number of two-qubit gates in $C$ that act on the qubits $\{i,j\}$. This summary is the input to the assignment of the qubits: a larger $A_{ij}(C)$ indicates a stronger interaction between $i$ and $j$ and encourages them to be closer together under $B_{\mathrm{eff}}$.

\softpara{Time and Arrivals.}
Each circuit $C$ arrives at time $t_c$ generated by a homogeneous Poisson process with rate $\lambda>0$. Equivalently, if $t_1<t_2<\cdots$ are arrival times, the inter-arrival gaps $\Delta_k:=t_k-t_{k-1}$ are i.i.d.\ $\mathrm{Exp}(\lambda)$ with $\mathbb{E}[\Delta_k]=1/\lambda$, and the number of arrivals in $[0,t]$ is $\mathrm{Poisson}(\lambda t)$. Arrivals are \emph{exogenous} (independent of scheduling and execution). Upon arrival, $C$ joins the waiting pool until admitted. We record its start time $s_c$ at admission and its completion time $e_c$.

\para{Qubit-Allocation Function $\boldsymbol{\mu}$.}
At time $t$, the qubit-allocation function for circuit $C$ is denoted $\mu_t$\footnote{Notation: $\mu_t$ is the \emph{operator} (a function of the current state and the circuit $C$). When $C$ is admitted at time $s_c$, we write the realized placement as $\eta_c := \mu_{s_c}(C)$, which is an injective map $Q(C)\hookrightarrow V_M^{\mathrm{free}}(s_c)$. In formulas later, we sometimes write $\mu_t(C)$ when the output map itself is intended; otherwise, we use $\eta_c$ to emphasize the committed placement.}. It returns an \emph{injective} map
\[
\mu_t(C):\; Q(C)\hookrightarrow V_M^{\mathrm{free}}(t).
\]
Here $Q(C)$ is the set of qubits of $C$, and $V_M^{\mathrm{free}}(t)\subseteq V_M$ are the memories not reserved by the active circuits at time $t$. Feasibility requires $|Q(C)|\le |V_M^{\mathrm{free}}(t)|$; otherwise, $C$ is deferred. Once admitted, $\eta_c$ remains fixed until $C$ completes (no migration). For completeness, Section~\ref{sec:gen} also explores a limited dynamic reassignment (e.g., teleport-and-carry at batch boundaries) that preserves the decoherence window~$\tau$; unless stated otherwise, our baseline results use fixed placements.

\para{Online Scheduling and Distribution of Quantum Circuits. (\onlinedqc).}
Given the quantum network with congestion-aware effective distances and an exogenous arrival process of circuits with arrival times, decide \emph{online}, at each decision time $t$, the following for any waiting circuit $C$ to minimize the makespan subject to network/fidelity constraints:
\begin{itemize}
    \item \emph{Admission / Start Time:} Defer $C$ or admit with a start time $s_c\!\ge\!t_c$.
    \item \emph{Qubit Allocation:} If admitted at $s_c$, choose an injective mapping $\mu_{s_c}(C)$.
    \item \emph{Execution Scheme:} Construct a plan that schedules all remote EPs required by $C$ in $\mu_{s_C}(C)$, choosing routes and swapping trees with respect to $w_{\mathrm{eff}}$ (i.e., distances in $B_{\mathrm{eff}}$). The plan coordinates across all admitted circuits to respect channel capacities on shared links. EP scheduling follows the circuit’s gate order: each EP is consumed when its corresponding two-qubit gate executes, and its generation is timed to occur within the decoherence window. See~\cite{dqcqr} for details on the consumption–generation coupling.
\end{itemize}

\softpara{Objective.}
Our default objective is to minimize the makespan of all the circuits:
\[
\min\ \max_{c} e_c,
\]
and we also report the mean waiting time $\frac{1}{|\{C\}|}\sum_c (e_c - t_c)$ in the evaluation. All decisions are made online (without look-ahead), using only the information available up to each decision time $t$.

\softpara{Complexity.}
Even under simplified assumptions (single circuit, fixed $B_{\mathrm{eff}}$, infinite channels and $\tau=\infty$), the \emph{qubit-allocation} subproblem~\cite{dqcqr},
\begin{align*}
&\min_{\eta_C}\ 
\mathsf{cost}\bigl(\eta_C \mid A(C),\,B_{\mathrm{eff}}\bigr)\\
&\quad=\min_{\eta_C}\ 
\sum_{i<j} A_{ij}(C)\,
B_{\mathrm{eff}}\!\bigl(\eta_C(i),\,\eta_C(j)\bigr),
\end{align*}
is precisely the \emph{Quadratic Assignment Problem} (\qap), a (strongly) NP-hard problem; thus, NP-hardness already arises at the allocation step and persists in the online setting.

\section{\bf Related Work}
\label{sec:related}

\para{DQC Primitives and Partitioning (Single Circuit).}
Nonlocal gates across QPUs are realized via teleportation and cat-entangler/disentangler constructions, which enable distributing a monolithic circuit over multiple devices~\cite{Bennett+:93,YimsiriwattanaL:05}. Modern DQC compilers formalize distribution as communication minimization: hypergraph partitioning of the circuit interaction graph~\cite{Andres-MartinezH:19}, optimization of teleportation/ebit usage during distribution~\cite{zomorodi2018optimizing}, and circuit partitioning strategies tailored for distributed execution~\cite{Daei2020OptimizedQC}. These works primarily target \emph{communication cost} (e.g., remote-gate count) for a \emph{single known} circuit and typically assume on-demand remote operations with fixed costs.

\para{Latency-/Time-Aware DQC and Distributed Scheduling.}
Beyond ebit/min-cut style objectives, time-aware approaches explicitly incorporate execution latency under connectivity and remote-operation costs; e.g., formulations that couple mapping with a latency model to reduce overall circuit time~\cite{dqcqr}. Other lines consider distributed execution pipelines once a circuit is partitioned, proposing practical orchestration/scheduling mechanisms across devices~\cite{sundaram2023distributing,dqc-gen,g2021efficient}. Related systems work studies joint qubit placement and entanglement provisioning to reduce the completion time of a single DQC task; e.g., QuPEP~\cite{Zhan2024IWQoS} combines offline placement with online entanglement provisioning during task execution. 
At the quantum-network layer, Cicconetti et al.~\cite{Cicconetti2021TQE} study online scheduling of end-to-end entanglement requests under stochastic link availability, proposing FIFO-style variants for request selection.
Overall, the above efforts focus either on entanglement-request scheduling or on single-job/fixed-set execution; instead, we study online admission of circuits under shared memory contention with decoherence-aware EP batching.

\para{Positioning.}
Relative to the above, our work is an \emph{online} multi-circuit formulation: we handle streaming arrivals, make joint admission/placement decisions under finite memories, use congestion-aware effective distances arising from in-flight jobs, and enforce generation–consumption coupling so nonlocal resources are consumed before decohering. This bridges DQC compilation with queueing-style constraints that prior single-circuit or offline methods typically abstract away.

\section{\bf High-Level Approach}
\label{sec:high-level}

We tackle \onlinedqc with a single \emph{event-driven loop} executed at each decision time.

\para{Decision Time.}
A \emph{decision time} $t$ is the earliest instant at which (i) one or more new circuits arrive, or (ii) one or more running circuits complete. Events with the same timestamp are processed together.

\para{EP Batches.}
An \emph{EP batch} is a stage of EP generation in which tasks run concurrently; batches run sequentially and are chosen to satisfy the decoherence threshold $\tau$.

\para{Unified Loop at Time $\boldsymbol{t}$.}
Let $\mathcal{U}(t)$ be the unfinished EP tasks after closing any circuits that are completed at $t$, and let $\mathcal{W}(t)$ be the waiting pool (not-yet-admitted circuits with arrival time $\le t$).
\begin{enumerate}
  \item \emph{Snapshot State.} Snapshot the current network state $(G_M, B_{\mathrm{eff}})$; block memories already in use by $\mathcal{U}(t)$; rebuild $B_{\mathrm{eff}}$ to reflect the load of $\mathcal{U}(t)$ and determine the currently free memories.
  \item \emph{Select Bundle.} Initialize $S\!\leftarrow\!\emptyset$. Repeatedly add \emph{one} circuit from $\mathcal{W}(t)$ using the admission and packing algorithms of Section~\ref{sec:algo}, maintaining feasibility on the currently free resources and respecting $\tau$. Each add-candidate evaluation invokes the \emph{prediction oracle} below (Step~3), which returns a placement and a cached batch plan (hence a predicted completion time) under the current snapshot. Stop when no feasible or beneficial addition remains.
  \item \emph{Place and Schedule (Prediction Oracle).} For each $c\in S$, compute the placement of the qubits $\eta_c$ in the free memories with respect to $B_{\mathrm{eff}}$. Build the new EP demand $\mathcal{E}(S)$ and compute a batch sequence for $\mathcal{U}(t)\cup\mathcal{E}(S)$ starting at $t$.
  \item \emph{Commit Outputs.} For each admitted circuit, record its placement $\eta_c$ and predicted completion time. The cached batch plan is committed without recomputation; any work extending beyond $t$ becomes the updated unfinished set $\mathcal{U}(t^+)$; the next decision time is the earliest of the next arrival or the next completion.
\end{enumerate}

\section{\bf Online Admission and Packing Algorithms}
\label{sec:algo}

We follow the event-driven runtime in Section~\ref{sec:high-level}: at each decision time $t$, we repeatedly rank/evaluate candidates by invoking the prediction oracle (Step~3 of the Unified Loop) and then commit a chosen circuit by adopting its cached plan without recomputation.

\para{Prediction Oracle.}
At decision time $t$, the prediction oracle evaluates a candidate circuit and returns (i) a heuristic qubit placement $\eta$ for the QAP objective and (ii) a cached EP-batch plan for executing the candidate together with any currently unfinished EP work. The predicted completion time\footnote{Because EP latencies are stochastic, predicted completion times are approximate. However, decisions are re-made at every event time (arrival/completion), so errors do not accumulate across long horizons; the policies primarily rely on the \emph{relative} ranking of candidates under the same oracle, and the system corrects course as actual completions trigger the next decision.} (and tail when needed) is read from this cached plan under $\eta$. All oracle inputs are available online at time $t$ (e.g., topology/link parameters, the current unfinished-task set, and the currently free memories), and it assumes no knowledge of future arrivals or future stochastic outcomes.

\para{Algorithm 1. FIFO Admission (\bffifo).}
Admit circuits strictly by arrival time, provided that each circuit can start \emph{now}. The goal is predictable, low-overhead operation and fairness to arrival order. The steps are as follows:

\begin{enumerate}
\item 
\emph{Select Circuit.}
\begin{itemize}
\item Scan the waiting pool in arrival order.
\item Choose the first circuit whose qubit demand fits the currently free memories at $t$.
\end{itemize}

\item 
\emph{Commit.}
\begin{itemize}
\item Run a single prediction for the selected circuit at $t$ to obtain its cached plan and finish time.
\item Commit to the network using the cached plan.
\end{itemize}

\item 
\emph{Iterate.}
\begin{itemize}
\item Refresh the snapshot.
\item Repeat until no arrival-ordered circuit is feasible.
\end{itemize}
\end{enumerate}

\para{Algorithm 2. Shortest-Time Packing (\bfshortest).}
Greedily favor circuits that \emph{finish soonest under current congestion}. This targets high instantaneous throughput and reduces head-of-line blocking by launching “easy’’ jobs first. The steps are as follows:

\begin{enumerate}
\item 
\emph{Pick Candidates.}
\begin{itemize}
\item Gather all waiting circuits that fit the currently free memories at $t$.
\end{itemize}

\item 
\emph{Select Circuit.}
\begin{itemize}
\item For each candidate circuit:
\begin{itemize}
\item Run one prediction at $t$ to obtain its predicted latency.
\end{itemize}
\item Select the circuit with the smallest latency\footnote{To prevent starvation under sustained load, one can add a max-wait override: If the waiting time of any circuit exceeds $W_{\max}$, admit it by arrival order regardless of the predicted finish time; our evaluation uses finite workloads and did not require this safeguard.}; break the ties by earlier arrival.
\end{itemize}

\item 
\emph{Commit.}
\begin{itemize}
\item Adopt the cached plan of the selected circuit.
\end{itemize}

\item 
\emph{Iterate.}
\begin{itemize}
\item Refresh the snapshot.
\item Repeat until no feasible candidate.
\end{itemize}
\end{enumerate}

\para{Algorithm 3. Greedy Min–Marginal Finishing Time (\bfmmft).}
Minimize the \emph{one-step increase in the current makespan}. \mmft compares the marginal effect of each candidate on the remaining completion time of active circuits and admits the circuit that least extends the tail of the system, balancing throughput with tail-latency control and avoiding unnecessary congestion spikes. The steps are as follows:

\begin{enumerate}
\item 
\emph{Pick Candidates.}
\begin{itemize}
\item Gather all waiting circuits that fit the currently free memories at $t$.
\end{itemize}

\item 
\emph{Select Circuit.}
\begin{itemize}
\item Establish the baseline tail $T_n$: the remaining time until all currently running circuits are finished.
\item For each candidate:
\begin{itemize}
\item Predict the resulting tail $T_c$ if it starts now in the same snapshot $t$.
\item Compute its marginal increase $T_c - T_n$.
\end{itemize}
\item Choose the circuit with the smallest marginal increase; break the ties by earlier arrival.
\end{itemize}

\item 
\emph{Commit.}
\begin{itemize}
\item Adopt the cached plan of the selected circuit.
\end{itemize}

\item 
\emph{Iterate.}
\begin{itemize}
\item Refresh the snapshot.
\item Repeat until no feasible candidate.
\end{itemize}
\end{enumerate}

\para{Oracle Overhead and Scalability.}
The policies differ mainly in how many oracle evaluations they perform per decision time: \fifo evaluates at most the first feasible arrival-ordered candidate, whereas \shortest/\mmft may evaluate multiple feasible candidates at the same $t$ to obtain predicted completion times (and \mmft marginal tails). If $K(t)$ candidates are evaluated at time $t$, the oracle overhead scales linearly with $K(t)$ and can be bounded by limiting the candidate set (e.g., an arrival-ordered prefix) and reusing cached predictions within the same decision time.
\section{\bf Event-Time Generalizations for Online DQC}
\label{sec:gen}

We present two event-time generalizations that (i) leverage qubit teleportation to adjust placements during execution and (ii) optionally defer launches when free capacity is scarce and dispersed across many QCs, favoring compact placements over fragmented ones. Both operate locally at the decision time $t$ and integrate with the same event-driven loop and EP batching used throughout.

\para{Teleport-and-Carry.}
Allow limited, well-scoped relocation of qubits via teleportation at decision times to relieve hotspots, shorten effective paths in $B_{\mathrm{eff}}$, and consolidate activity where helpful, while keeping already-issued work intact.

\softpara{Action at $t$.} Consider moves only at EP batch boundaries or when a clear benefit is predicted (e.g., lower tail or freed bottleneck). Select a small set of qubits with the highest gain-to-cost ratio, teleport them to nearby free memories (prefer intra-QC or short-hop), and update the placement $\eta$ and the future EP demand for gates that have not yet been generated.

\para{Fragmentation-Aware Admission.}
Aggressive packing into many small, scattered free slots increases fragmentation and later contention. At time $t$, if free memories are scarce and spread over many QCs, we may defer the launch even when a circuit technically fits, favoring placements that use fewer QCs and larger contiguous blocks.

\softpara{Action at $t$.} Summarize the free capacity by QC (e.g., largest block per QC, number of QCs with free slots, dispersion across QCs). Admit only if the newcomer can be placed with a small QC footprint or sufficiently large per-QC blocks; otherwise wait for the next event so completions can consolidate capacity. Prefer launches that preserve large blocks and avoid cross-QC stitching.

\section{\bf Evaluation}
\label{sec:eval}

\begin{figure*}[t]
\vspace*{-0.1in}
    \centering
    \includegraphics[width=\textwidth]{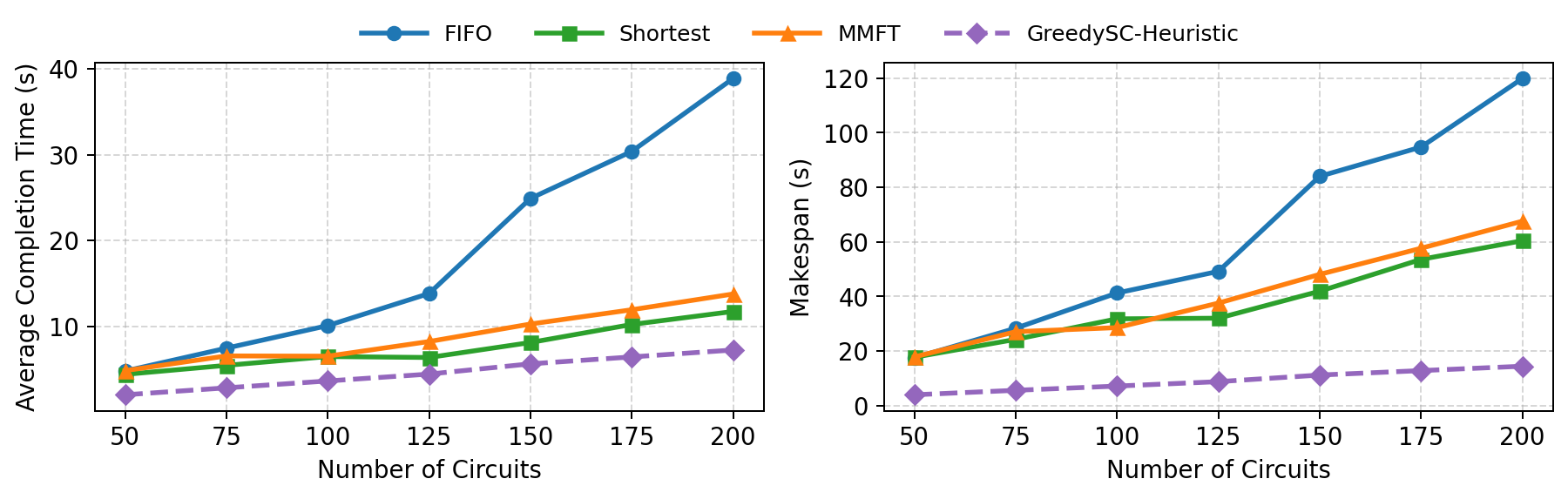}
    \caption{The total number of circuits per stream is varied. Panels show (a) average completion time and (b) makespan. Arrivals followed a Poisson process and all other network and runtime parameters were held fixed. Curves compared \fifo, \shortest, and \mmft within the same event-driven loop for the online case. \GSCH was included as the offline baseline for comparison.}
    \vspace*{-0.1in}
    \label{fig:cir_num}
\end{figure*}

\begin{figure*}[t]
    \centering
    \includegraphics[width=\textwidth]{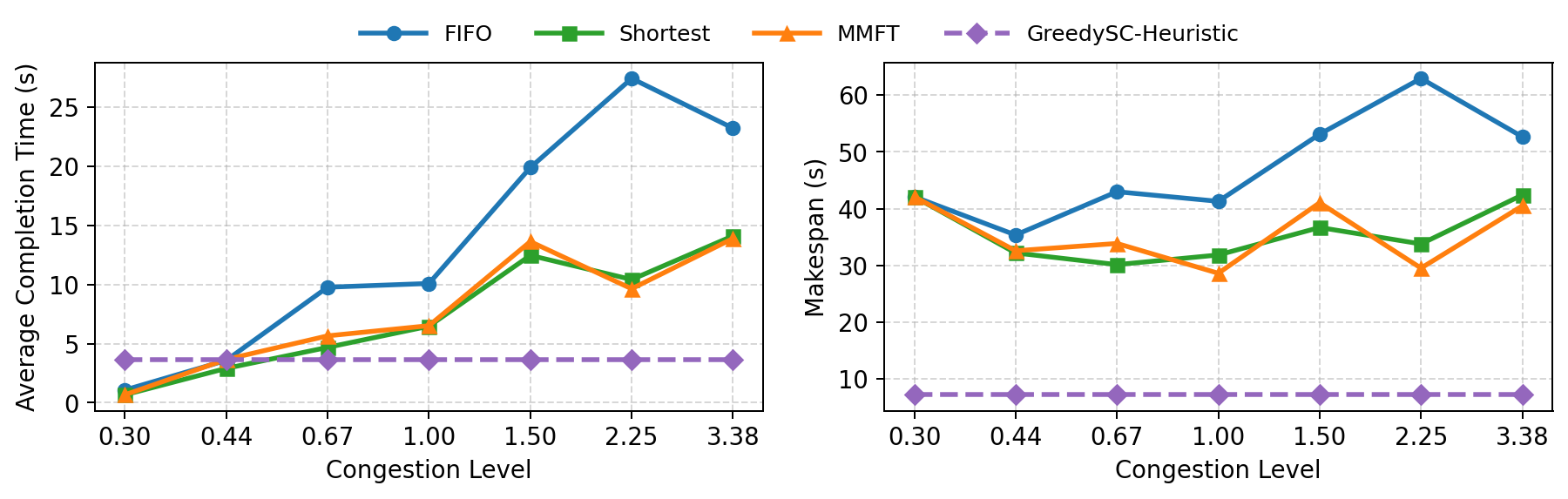}
    \caption{The offered load is varied by scaling the arrival rate (reported as a congestion level on the x-axis; higher values indicate heavier load). Panels show (a) average completion time and (b) makespan. Curves compare \fifo, \shortest, and \mmft for the online case, with \GSCH included as the offline baseline for comparison; the arrival rate does not affect the offline case.}
    \vspace*{-0.1in}
    \label{fig:time}
\end{figure*}

\begin{figure*}[t]
\vspace*{-0.1in}
    \centering
    \includegraphics[width=\textwidth]{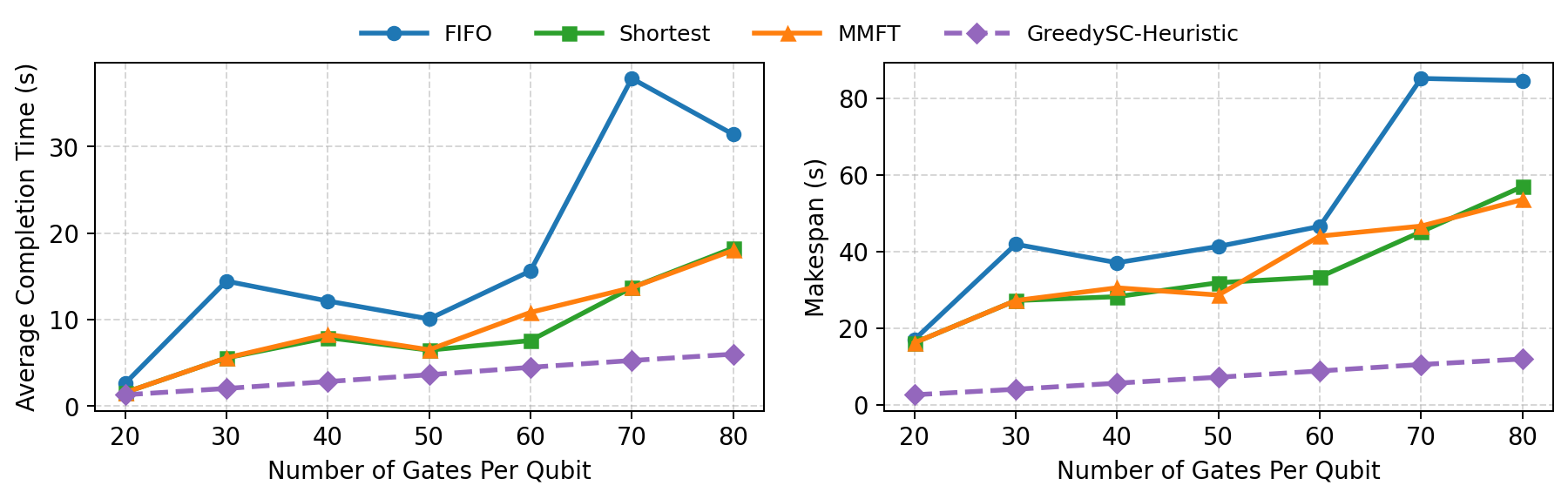}
    \caption{The number of gates per qubit is varied. Panels show (a) average completion time and (b) makespan. Each point aggregated results from circuits constructed at the indicated gate count under identical network and arrival settings. Curves compared \fifo, \shortest, and \mmft for the online case, with \GSCH included as the offline baseline for comparison.}
    \vspace*{-0.1in}
    \label{fig:gate}
\end{figure*}

In this section, we evaluate the \onlinedqc\ scheduler developed in Secs.~\ref{sec:high-level}–\ref{sec:algo} under the full execution model: congestion-aware distances, finite memories, stochastic EP generation with batching and a decoherence threshold. We compare admission/launch policies that share the same arrival stream of circuits and the same EP-generation back end, and we focus on completion-centric metrics (mean completion time and makespan). We use NetSquid~\cite{netsquid} as the quantum-network simulator to capture probabilistic link behavior, swapping across paths, channel concurrency, and qubit decoherence.

\para{Generating Random Network.}
The network has $10$ quantum computers, each with $10$ quantum memories (total $100$ memories).
We generate a Waxman host graph~\cite{waxman} by placing the computers uniformly at random in a $100\,\mathrm{km}\times100\,\mathrm{km}$ square and creating distance-biased links.
\eat{We then fix this topology and reuse it across all experiments and policies.}

\softpara{Network Parameters.}
Following~\cite{DPpaper}, we set: atomic-BSM success probability $0.4$ with latency $10\,\mu\mathrm{s}$; optical-BSM success probability $0.3$; atom–photon pair generation time $50\,\mu\mathrm{s}$ with success probability $0.33$; and a decoherence threshold of $1\,\mathrm{s}$.

\para{Generating Random Circuits.}
For the incoming circuits, we generate random quantum circuits with four knobs: number of circuits (default 100), qubits per circuit (uniform in 20--30), gates per qubit (default 50), and the fraction of two-qubit gates (fixed at 0.5).
Each circuit is built gate by gate: at each step, the gate type is chosen at random. Operands are then selected uniformly (two distinct qubits for two-qubit gates; one qubit otherwise).
For evaluation, we balance the size distribution by giving equal probability to each qubit count in 20--30, and we reuse the identical circuit set across all policies.

\softpara{Choosing the Poisson Arrival Rate.}
We set the arrival rate $\lambda$ in three steps.
\begin{enumerate}
    \item \emph{Probe Time.} Run a randomly sampled 20-qubit circuit and a 30-qubit circuit on the network and average their completion times; denote this average by $T_p$.
    \item \emph{Baseline Rate.} Set the default Poisson rate to
    \[
      \lambda_0 \;=\; \frac{1}{0.75\,T_p},
    \]
    i.e., the mean inter-arrival gap is $0.75\,T_p$.
    \item \emph{Arrival Intensity Control.} Introduce an arrival rate multiplier $r>0$—also shown as the \emph{congestion level}—to vary traffic:
    \[
      \lambda \;=\; r\,\lambda_0
      \qquad\text{(equivalently, mean gap }=0.75\,T_p/r\text{).}
    \]
 The larger $r$ means more traffic and more congestion.
\end{enumerate}

\para{Algorithms Compared.}
The evaluation compared three \emph{online} admission/launch policies—\fifo, \shortest, and \mmft—together with an \emph{offline} baseline, \GSCH~\cite{Yang2025Efficient}. The online policies use a common snapshot and prediction oracle at each decision time, whereas the offline baseline has full lookahead over the entire workload and is therefore insensitive to the arrival rate. \GSCH\ is included because, in that previous work, it provided the best performance among \emph{polynomial-time} batching schedulers; the offline optimal scheduler there is exponential-time and thus not suitable as a per-decision baseline in our online setting.

\softpara{Metrics.}
For online policies, the metrics reported are the average completion time $\frac{1}{n}\sum_{c=1}^{n}(e_c - t_c)$ and makespan $\max_{1\le c\le n} e_c$, where $t_c$ and $e_c$ are the arrival and completion times of the circuit~$C$, and $n$ is the total number of circuits. For the offline baseline, the average completion time is measured analogously with a common release time (all $t_c \equiv 0$), i.e., $\frac{1}{n}\sum_{c=1}^{n} e_c$; consequently, the chosen arrival rate (congestion level) does not affect the offline metric.

\para{Evaluation Results.}
Each experiment varied one parameter at a time, while others remained at their defaults. We observe the following:
\begin{enumerate}
\item
\softpara{Circuit Number.}
Both metrics increase with circuit number. \shortest is consistently best across the plotted settings, while \fifo is worst under growing contention. \GSCH (offline) remains below all online curves, with a widening gap at a larger circuit number.
\item
\softpara{Congestion Level.}
Both metrics increase with arrival rate. \shortest and \mmft perform similarly, with \mmft showing larger fluctuations\footnote{\mmft can be more volatile because it optimizes a \emph{marginal} predicted tail: when the system is near saturation, small snapshot changes (or small prediction differences between candidates) can flip the marginal comparison, leading to policy churn and higher variance even when mean performance is comparable.}; \fifo is worst. \GSCH is insensitive to the rate. At low congestion, \GSCH may slightly underperform the online policies because its global batching can introduce avoidable waiting, whereas the online policies can admit and finish circuits sequentially with little queueing.
\item
\softpara{Gate Number.}
Both metrics increase with gate number. \shortest and \mmft are comparable; \fifo is the worst. \GSCH remains the best across the sweep.
\item
\softpara{Qualitative Robustness.}
Across all sweeps, the qualitative ordering was consistent. \GSCH\ (offline) generally formed a lower envelope; among the online policies, \shortest and \mmft were close-\shortest typically achieved the best average completion time (often best overall); \fifo consistently underperformed\footnote{\fifo remains attractive in operational regimes where scheduler overhead must be minimal or strict arrival-order fairness is required, even if its completion-time metrics are worse under moderate/heavy contention.}. Gaps widened with increasing congestion, larger circuit numbers, and deeper circuits. At very low congestion, \GSCH’s global batching could be slightly worse than the online policies’ sequential admission.
\end{enumerate}

\section{\bf Conclusion}
\label{sec:conclusion}

We studied \onlinedqc on the quantum network and presented an event–driven framework that maintains \emph{congestion-aware routing} via $B_{\mathrm{eff}}$, uses it to place qubits, and to batch EP generation on-demand under a decoherence window $\tau$. We instantiated three online admissions—\fifo, \shortest, and \mmft—and compared them against an offline global scheduler (\GSCH). Across circuit, contention, and gate sweeps, \fifo underperformed; \shortest was often the best among the online algorithms. Overall, per-event prediction coupled with decoherence-aware batching yields robust online performance without lookahead.

\begin{small}
\bibliographystyle{IEEEtran}
\bibliography{ref}
\end{small}
\end{document}